\documentclass[12pt,a4paper]{article}
\usepackage{epsfig}
\def\gevs{{\rm\, GeV}^2}

\def\beq{\begin{equation}}
\def\eeq{\end{equation}}
\def\bea{\begin{eqnarray}}
\def\eea{\end{eqnarray}}

\def\({\left(}   
\def\){\right)}   

\def\eq#1{{Eq.~(\ref{#1})}}

\def\npb#1#2#3{    {\it Nucl. Phys. }{\bf B#1} (19#2) #3}
\def\plb#1#2#3{    {\it Phys. Lett. }{\bf B#1} (19#2) #3}
\def\prd#1#2#3{    {\it Phys. Rev. }{\bf D#1} (19#2) #3}
\def\prep#1#2#3{   {\it Phys. Rep. }{\bf #1} (19#2) #3}

\def\sjnp#1#2#3{   {\it Sov. J. Nucl. Phys. }{\bf #1} (19#2) #3}

\def\epj#1#2#3{    {\it Eur. Phys. J. }{\bf C#1} (19#2) #3}

\newcommand{\slope}{\frac{\partial F_2}{\partial \ln Q^2}}

\begin
{document}

\vspace{1cm}

\title { {\LARGE\bf{SCALING VIOLATIONS IN THE $\mathbf{Q^2}$}}\\[2ex]    
{\LARGE\bf{ LOGARITHMIC DERIVATIVE OF $\mathbf{F_2}$ }} }
\author{
{\bf
E.~Gotsman\thanks{E-mail: gotsman@post.tau.ac.il}~$\,^{a}$,
\quad E. ~Ferreira \thanks{E-mail: erasmo@if.ufrj.br}~$\,^{b}$,
\quad  E.~Levin\thanks{E-mail: leving@post.tau.ac.il,
 levin@mail.desy.de}~$\, ^{a}$,
}\\
{\bf 
 U.~Maor\thanks{E-mail: maor@post.tau.ac.il}~$\,^{a}$
 \quad and \quad
E. Naftali\thanks{E-mail: erann@post.tau.ac.il}~$\,^{a}$ 
}\\[10mm]
 {\it\normalsize $^a$HEP Department}\\
 {\it\normalsize School of Physics and Astronomy,}\\ 
 {\it\normalsize Raymond and Beverly Sackler Faculty of Exact Science,}\\
 {\it\normalsize Tel-Aviv University, Ramat Aviv, 69978, Israel}\\[0.5cm]
{\it\normalsize $^b$  Instituto de Fisica, Universidade Federal do 
Rio de Janeiro}\\
{\it \normalsize Rio de Janeiro RJ21945-970, BRASIL}
}
\maketitle
\thispagestyle{empty}
\begin{flushright}
\vspace{-14cm}
{ TAUP 2653-2000}\\
{ \today}\\
\end{flushright}

~
\vspace{12cm}

\begin{abstract}
We examine the latest HERA experimental measurements 
of the $Q^2$ logarithmic derivative of the proton
structure function.
We analyze the characteristics of  DGLAP with and without screening, as
well as a  Regge type model, and compare their predictions to
all available data on $\slope$
including those for fixed W. Our results show that the present data 
can be described in pQCD taking into account shadowing
corrections. However, such a description cannot be considered 
a conclusive   signal of gluon saturation  since the
experimental data on $\slope$
do not allow one to discriminate between the various  approaches ,
in spite of their very  different construction.
\end{abstract}

\thispagestyle{empty}

\newpage
\setcounter{page}{1}

\section{Introduction}

The physics of small $Q^2$ and small $x$ is
associated with
the search for the scale where gluon saturation \cite{MUEsat},
implied by s-channel unitarity \cite{SAT}, sets in.
Gluon saturation signals the 
transition from the perturbative (hard) to the non perturbative 
(soft) QCD regime.
We expect this transition to be  preceded by signatures of
screening corrections (SC)
which  should  be
experimentally visible even though the relevant scattering amplitude has
not yet reached the unitarity (black disk) limit.
Although, the general theoretical framework for saturation is reasonably
well understood \cite{MUEsat,SAT}, the specific kinematic domain
where we
expect to see evidence of the role it plays, is as yet not determined.  
Consequently, this  QCD component depends on the relevant DIS
experimental data and the associated phenomenology.
While the global analysis of the
proton structure function $F_2$
(or $\sigma_{tot}^{\gamma^*p}(W,Q^2)$) data shows no significant 
deviations
from DGLAP \cite{DGLAP2}, there are dedicated HERA measurements of the 
$Q^2$ logarithmic derivative of the proton structure function $F_2$
\cite{caldwell,H1slope,ZEUSslope}, which
add to our knowledge as they provide
detailed information on the rate of change of the logarithmic $Q^2$
dependence of $F_2$, and, hence, they magnify possible deviations from the
expected DGLAP behaviour at small x and small $Q^2$.

The interest in ($\slope$)   stems from the
observation that: 
\begin{enumerate}
\item\,\,\,In leading order  DGLAP evolution the $Q^2$ logarithmic slope
of $F_2$  at low 
$x$,   is
directly proportional to the gluon structure function
\cite{PryB}, since 
\beq\label{A}
\frac{\partial F_2(x,Q^2)}{\partial \ln Q^2}\,=
\,\frac{2\alpha_S}{9\pi}xG^{DGLAP}(x,Q^2),
\eeq
where $xG^{DGLAP}(x,Q^2)$ denotes the DGLAP gluon distribution of the 
proton.
\item\,\,\, $\slope$ is also related to the 
dipole cross
section \cite{DIPOLE}
\beq \label{AA}
\frac{\partial F_2(x,Q^2)}{\partial \ln Q^2}\,=\, \frac{Q^2}{3  \pi^3}
\sigma_{dipole} (r^2_{\perp}, x) \,=\, \frac{Q^2}{3 \pi^2}\int \,d b^2
Im a_{el}( r^2_{\perp}, x; b);
\eeq
where $ a_{el}( r^2_{\perp}, x)$ is the elastic scattering amplitude at
fixed impact parameter of a dipole of  size $ r_{\perp} = 2/Q$  with
energy $x$. For the  amplitude $ a_{el}( r^2_{\perp}, x)$ we have a
non-linear equation in the region of high density QCD  (see
Refs.\cite{SAT,EQ}) as well as the unitarity constraint  $a_{el}(
r^2_{\perp}, x; b) \,\leq\,1$.  Using \eq{AA} one can  extract the 
dipole - target amplitude directly from  the experimental data.

\end{enumerate}

In this letter we present a detailed study of
$\frac{\partial F_2(x,Q^2)}{\partial \ln Q^2}$.
Our goal  is to utilize this quantity, which is relatively well measured,
to examine whether   the system of partons has reached a  QCD regime of
gluon saturation. The complexity and difficulty of this investigation is
illustrated by Fig.~1 which shows two different limits of $\slope$: the
DGLAP limit  at low $x$  for  different parton distribution functions 
(pdf) \cite{GRV98,MRS99,CTEQ5} solutions of  the evolution
equations; and the unitarity limit at 
very long distances (low $Q^2$). From  Fig.~1 one can see that (i)
both the DGLAP equation and gluon saturation which leads to the
unitarity bound displayed in Fig.~1 have the same qualitative behaviour
decreasing  at low $Q^2$; and (ii) the saturation scale ($Q_s(x)$) at
which the
DGLAP solution crosses the unitarity bound,  differs for different
parameterizations.  $Q_s(x) \approx  4 - 5 \gevs$   for GRV98
and   $Q_s(x) \approx  1 - 2 \gevs$ for CTEQ5 and MRS99 parameterizations.
Such a low value for the saturation scale indicates that we shall have 
difficulty attempting to distinguish between models with 
gluon saturation effects, and 
models that include  soft and hard interactions, since the matching
of the components occurs
at similar values of $Q^2$ ($ \approx 1\,GeV^2$).

\begin{figure} \centerline{\epsfig{file=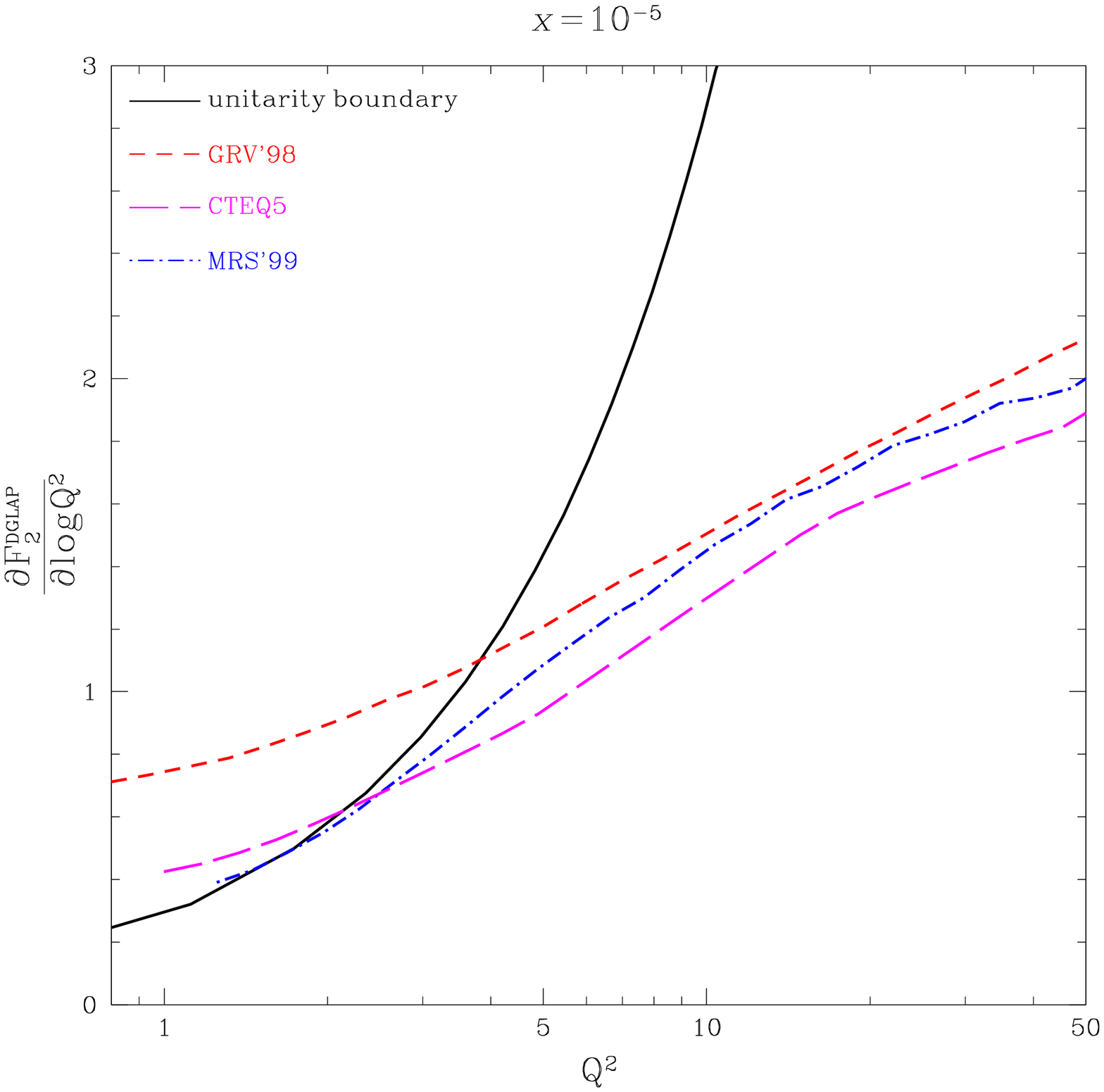,width=100mm}}
\caption{\it  The $Q^2$ behaviour of $\slope$ at $x = 10^{-5}$ ($Q^2$ in 
$GeV^2$).}
\label{Fig.0}
\end{figure} 

In our study we will incorporate the following initial observations and
constraints:

\begin{enumerate}
\item\,\,\,Recent measurements of $F_{2}$ \cite{ZEUSlowQ2}, extending to
very low
$Q^2$, find that it has a monotonic behaviour down  to the
photo production limit. These data suggest a smooth transition from the
perturbative (DGLAP, high $Q^2$)
to the non perturbative (Regge or classical gluon
field, exceedingly small $Q^2$) regime.
\item\,\,\,Preliminary HERA data \cite{H1slope,ZEUSslope} on
$\slope $ at small $Q^2$ and small
$x$ do not show any turnover when plotted at fixed $x$ or fixed $Q^2$.
\item\,\,\,An approach which provides a way of successfully describing 
$\slope$ 
should  also be able to reproduce data in other channels which
are sensitive to $xG(x,Q^2)$ such as $F_{2}^c$, $F_{L}$ and the photo and
DIS production of heavy vector mesons.

\end{enumerate}

The main goal of this letter is to investigate and compare:

\begin{enumerate}
\item\,\,\,The ability of conventional DGLAP, using the pdfs
\cite{GRV98,MRS99,CTEQ5}, to reproduce
the available data on $\slope$.
\item\,\,\,The role of screening corrections (SC). In this approach
\cite{GLMslope,GLMNslope,AMIRIM}, DGLAP
is modified
by SC due to the $q \bar{q}$ percolation through the target, as well as
by
modifications due to screening in the gluon distribution $xG(x,Q^2)$. 
\item\,\,\,Within the framework of pQCD no allowance is made for explicit soft
contributions, both in
the non-screened and screened formulations of DGLAP. The latter
provides a smooth interpolation between the high and low $Q^2$ regimes. 
In the present
investigation we have checked the validity of our methods at low values of
$Q^2$, i.e. just above the DGLAP $Q_0^2$ threshold (defined for each pdf
input). For the case with  SC, we have also considered an
extrapolation for $Q^2 \,\leq\, Q_0^2$. Our aim is to determine the values
of $Q^2$ that one can successfully describe without the help
a soft component. To this end we compare with the
Donnachie and Landshoff (DL) two Pomeron model \cite{DL2P}, which
explicitly uses the sum of a hard and a soft Pomeron.
\item\,\,\,We conclude with a critical discussion and caution regarding attempts
that
have been made to determine the gluon saturation scale using data with
highly correlated variables such as in the Caldwell and fixed W plots
\cite{caldwell,ZEUSslope,G-BW}. We also comment on
the
general problems associated with gluon saturation and screening
corrections relating to other  models which include SC
\cite{G-BW,KMP,CKMT,Forshaw}.

\end{enumerate}

\section{The small $Q^2$ and small $x$ behavior of 
$\slope$ in  various models }

As is well known, a global
DGLAP analysis of the data with the recent pdfs   
\cite{GRV98,MRS99,CTEQ5} is in good agreement
with the experimental data. A study \cite{ADL}
comparing the screened and non screened DGLAP calculations of $F_2(x,Q^2)$ 
showed only a small difference due to SC even in the small $Q^2$ and $x$
region attained by
present HERA measurements.
A significant deviation of the data from \eq{A},
where $xG^{DGLAP}$ is obtained from the global $F_2$ analysis,   
may serve as an experimental signature indicating the growing 
importance of unitarity corrections. 
This was recently suggested by Caldwell \cite{caldwell}, showing a rather  
complicated plot of $\slope$ in which each
measured  point
had different correlated values of $Q^2$ and $x$.
The Caldwell plot suggested a dramatic turnover of 
$\slope$ corresponding to $Q^2$ of about $3 \,\, GeV^2$ 
and $x<5 \cdot 10^{-3}$, in contrast to the behavior expected 
from GRV94 \cite{GRV94} at sufficiently small $Q^2$ and $x$. The
problem with this presentation is that, as suggestive as it may seem, it
does not discriminate between different dynamical interpretations 
\cite{GLMslope,GLMNslope,DL2P,G-BW,KMP,Forshaw}.
It is, actually compatible with an overall data generator \cite{ALLM} as
well as the latest pdfs  which were adjusted to
account for this observation.

Better discrimination is achieved if we carefully
study the small $Q^2$ and $x$
dependences of $\slope$ at either fixed $Q^2$ 
or fixed $x$ values, so as to be free of the kinematic correlation
between
$Q^2$ and $x$ that is endemic in the data displayed in  the Caldwell
plot. Such preliminary HERA data 
have recently became available \cite{H1slope,ZEUSslope}. Note that
the method that has been used to specify the logarithmic derivative
of the structure function $F_2$ in \cite{H1slope,ZEUSslope} is by
fitting the $F_2$ data at fixed $x$, to the expression
\begin{equation}\label{abc1}
F_2(x,Q^2) = a(x) +  b(x) \ln Q^2 + c(x)[\ln Q^2]^2
\end{equation}
and then make the identification
\begin{equation}\label{abc2}
\partial F_2 / \partial \ln Q^2 = b(x) + 2c(x) \ln Q^2
\end{equation}    
H1 \cite{H1slope}  compared the results  using Eq.~(\ref{abc2}) with the
results  obtained by taking the  derivative of their QCD  
parameterization (H1 QCD fit). They find that for  $x \, \geq \, 0.0002 $,
the L.H. and R.H. sides of Eq.~(\ref{abc2})  match very well, however, for 
$x \, < 2\times 10^{-4}$
there is a discrepancy. We note that the experimental data of
\cite{H1slope,ZEUSslope} are correlated,
such that the lower values of $x$ are
associated with the lower values of $Q^2$.

At the initial stage of our investigation we examined whether DGLAP
(using the pfds \cite{GRV98,MRS99,CTEQ5} in the NLO
approximation) reproduce the 
data. The results are shown in Fig.~2. We note that:

\begin{enumerate}
\item\,\,\,GRV98 overestimates the low $Q^2 \,\leq\, 5\, GeV^2 $ data.
For higher $Q^2$, the reproduction of the data is good.
\item\,\,\,MRS99 underestimates the data at $Q^2$ = 1.9 and 40 $GeV^2$. 
It is above
the data in the $ 3\, \leq \,Q^2\, \leq 10\, GeV^2$ range and does well
for $Q^2$ = 12 and 30 $GeV^2$.
\item\,\,\,CTEQ5 (we have used CTEQ5HQ as it reproduces 
the energy dependence
of $J/\psi$ photo production \cite{GFLMNpsi})
provides a good reproduction of the experimental data over the entire
range $1.9\, \leq \,Q^2\, \leq 40\, GeV^2$. 
A few words of explanation are called for regarding the numerical NLO
calculation of
$F_2$ from the CTEQ pdf (we discuss only CTEQ in this context since
both GRV98 and MRS99 supply a code for calculating the structure
function, from which the parton densities have been
parameterized). When calculating $F_2$ in NLO one should take care
when inserting the threshold 
for  charm production, so that $F_2$ remains a smooth
function of $Q^2$. The effect of the threshold is small if one looks
at the structure function, but it can be rather large when examining
it's $Q^2$ derivative. We found that by following the consistent treatment
of charm in \cite{MRRS}, we obtain a smooth behaviour  of both $F_2$
and  $\slope$.

\end{enumerate}.

In order to study the role of SC in our calculations, we follow
the eikonal SC formalism presented in Ref.\cite{GLMNslope}\footnote{For
theoretical and phenomenological status of this approach see
Ref. \cite{AMIRIM}.}, where
screening is calculated in DLA in both the quark sector, to account for
the
percolation of a $q\bar q$ through the target, and the gluon sector, to
account for the screening of $xG(x,Q^2)$. The factorizable result that we
obtain is
\beq\label{B}
\frac{\partial F_2^{SC}(x,Q^2)}{\partial \ln Q^2}\,=
\,D_q(x,Q^2)D_g(x,Q^2)
\frac{\partial F_2^{DGLAP}(x,Q^2)}{\partial \ln Q^2}.
\eeq
SC in the quark sector are given by
\beq\label{C1}
D_q(x,Q^2)\frac{\partial F_2^{DGLAP}(x,Q^2)}{\partial \ln Q^2}\,=
\,\frac{Q^2}{3\pi^2} \int\,db^2\(1\,-\,e^{-\kappa_q(x,Q^2;b^2)}\),
\eeq
\beq\label{C2}
\kappa_q\,=\,\frac{2\pi \alpha_S}{3Q^2}xG^{DGLAP}(x,Q^2)\Gamma(b^2).
\eeq
The calculation is  simplified if we assume a Gaussian 
parameterization for the two gluon non perturbative form factor,
\beq\label{C3}
\Gamma(b^2)\,=\,\frac{1}{R^2}e^{-b^2/R^2}.
\eeq
SC in the gluon sector are given by 
\beq\label{D1}
xG^{SC}(x,Q^2)\,=\,D_g(x,Q^2) xG^{DGLAP}(x,Q^2),
\eeq
where
\beq\label{D2}
xG^{SC}(x,Q^2)\,=\,
\frac{2}{\pi^2}\,\int_{x}^{1}\,\frac{dx^{\prime}}{x^{\prime}}
\int_{0}^{Q^2}\,dQ^{\prime\,2}\,
\int db^2\,\(1\,-\,e^{- \kappa_g(x^{\prime},Q^{\prime\,2};b^2)}\),
\eeq
and $\kappa_g=\frac{9}{4}\kappa_q$.

An obvious difficulty in the above calculation of $xG^{SC}$ stems
from the
fact that the $Q^{\prime\,2}$ integration spans not only the short (pQCD),
but also the long (npQCD) distances. To overcome this difficulty we assume
that
\beq\label{D3}
xG\(x,Q^2\,<\,\mu^2\)\,=\,\frac{Q^2}{\mu^2}xG\(x,\mu^2\).
\eeq
Our choice of the above extrapolation is
motivated by the gauge invariance requirement 
that $xG\,\propto\,Q^2$ when $Q^2\,\rightarrow\,0$, and is supported
by recent ZEUS measurements of $F_2$ at exceedingly small $Q^2$
\cite{ZEUSlowQ2}.

The SC calculation described above can be applied to any given input pdf
where the only adjusted parameters are $R^2$ and $\mu^2$. For the
radius  $R^2$, see \eq{C3}, we use the value  8.5 $GeV^{-2}$
which is determined directly from the
forward slope of $J/\Psi$ photo production \cite{MRRS}. 
$\mu^2$ is conveniently fixed at $Q_0^2$, the lowest $Q^2$ value
accessible for  the input pdf we use. Once we have chosen our pdf,
our SC
calculation is essentially parameter free.  
We have checked that our output results are not sensitive to the
fine tuning of these fixed parameters.

Before presenting our results we would like to recall two important
features of SC: 

\begin{enumerate}
\item\,\,\,They can only dampen (reduce) the results obtained with the
unscreened input dpfs. 
\item\,\,\,They are negligible for large enough $Q^2$ and/or $x$. 
This can be
easily deduced from the functional dependence of $\kappa_{q}$ and
$\kappa_{g}$ on $Q^2$ and $x$ \cite{GLMNslope}. 

\end{enumerate}

Throughout the SC calculation we have used as input the
$\overline
{MS}$ version of NLO GRV98. The results obtained using GRV98 DIS are
very similar. 

Our fixed $Q^2$, and fixed $x$ results are displayed in
Fig.~3 and Fig.~4, respectively.
As can be seen in the limit of small $Q^2$ and $x$ there is a
significant difference between the screened and non screened values
of $\slope$. As expected the SC results are
smaller and softer than the non screened input. All in all, 
our reproduction of the experimental data for $Q^2\, > \,2\,GeV^2$
is very good.    
The ZEUS fixed $Q^2 \, $ = 1.9 $GeV^2$ data are somewhat softer than
our predictions. This feature is conspicuous when we compare with the
very small $Q^2$ = 0.3 and 0.75 $GeV^2$ data, where  
we have made use of the extrapolation given in Eq.(11). In Fig.~4 we see
that, using this extrapolation, screened GRV'98 is also successful in
describing the data at values of $Q^2 \,\leq\,1\,GeV^2$, where in the DL
parameterization (see Fig.~6) the soft component dominates.

To further examine the role of the soft (non perturbative) component in 
$\slope$, we compare the above DGLAP results with 
the DL two Pomeron model \cite{DL2P}. The DL parameterization is based
on the Regge
formalism, and consists of the coherent sum of contributions from a
hard and soft Pomeron, a normal Reggeon and an additional
contribution from the charmed sector which is proportional to the
hard Pomeron. Each of these fixed j-poles are multiplied by a fitted
$Q^2$ form factor.  The hard Pomeron with a fixed trajectory has an
intercept of 1.43 at t=0, while the soft Pomeron has an intercept of
1.08 at t=0.  As can be seen in Fig.~3, the DL model 
reproduces the data at all values of $Q^2$.

We conclude that one can obtain a good description of $\slope$
data with $Q^2\, \geq \, 1.9$ $GeV^2$ 
using any of the following options: 

\begin{enumerate}
\item\,\,\,Conventional DGLAP evolution using CTEQ5 pdfs as input. 
\item\,\,\,DGLAP evolution using GRV98 pdfs as input, provided SC are included. 
\item\,\,\,The two Pomeron (DL)  model combining hard and soft
components. 

\end{enumerate}

For smaller  $Q^2 \, (< \, Q_{0}^{2})$, a soft contribution is called for,
for which there are several parameterizations available, but not yet a
precise theory.

\begin{figure} \centerline{\epsfig{file=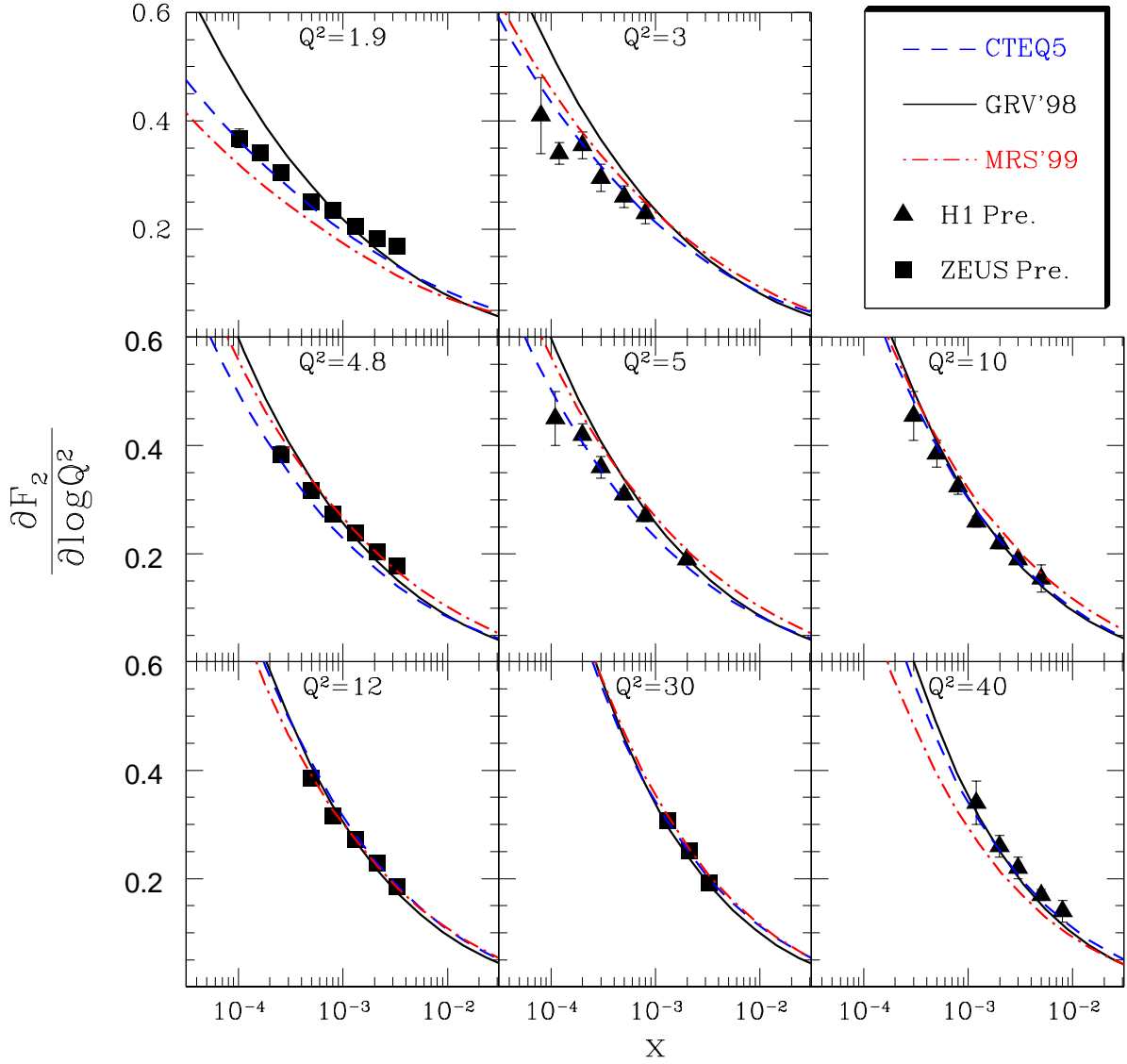,width=170mm}}
\caption{\it x dependence of HERA data for the logarithmic slope  at
fixed $Q^2$ (in $GeV^2$)
compared with  calculations for the unscreened pdfs.}
\label{Fig.1}
\end{figure}

\begin{figure} \centerline{\epsfig{file=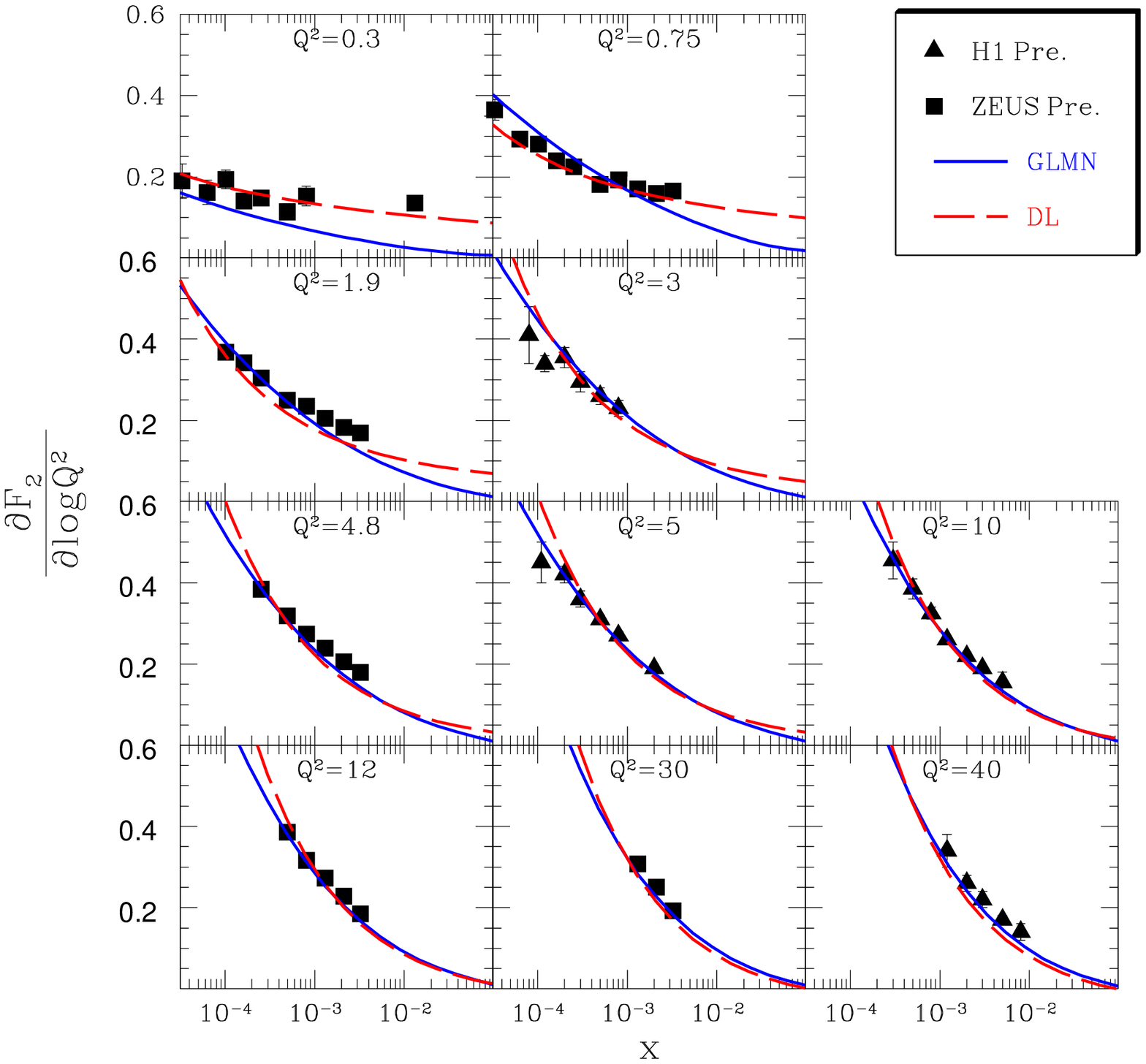,width=170mm}}
\caption{\it x dependence of HERA data for logarithmic slope  at fixed
$Q^2$ (in $GeV^2$)
compared with  our calculations for screened GRV98 and the DL model.}
\label{Fig.2}
\end{figure}

Recently, Golec-Biernat and Wuesthoff \cite{G-BW} have suggested
studying the
$Q^2$ and $x$ behaviour of $\slope$ at fixed W
as a method to determine the gluon saturation scale from the
anticipated turnover in these plots.   The new ZEUS
low $Q^2$ presentations \cite{ZEUSslope} of these plots show, indeed, the
anticipated turnover structure in these figures, suggesting that
gluon saturation is attained at $Q^2\,\simeq \,$ a few Ge$V^2$.  
We advocate using  the invariant variables  $x$ and
$Q^2$ when   describing  $F_2$.  Introducing other variables such as W to
study the structure
function gives rise to spurious effects which are predominantly
kinematic.  In the specific procedure suggested by Golec-Biernat and
Wuesthoff, the combination of the kinematic relation between $x, Q^2$
and W with the very general behaviour of $xG(x,Q^2)$ is sufficient to
produce a turnover. Its exact location depends on the details of the
numerical input. Consequently, the suggested fixed W plots do not
convey any new dynamical information on the saturation problem even if
such information is
contained in the analyzed data. We note that a fixed $W$ plot is natural
and informative for Regge type models.
It appears that any parameterization
for $F_2$ (or $xG$) which has factorizable $Q^2$ and $x$ dependences,
and provides a reasonable description of the data such as the
Buchmueller- Haidt (BH) model \cite{H}, gives rise to the fixed W 
turnover effect.

To illustrate this point we consider the fixed W behaviour in three
parameterizations  which have completely different dynamics:

\begin{enumerate}
\item\,\,\,DGLAP with CTEQ5 pdf. As noted, the 
results of this parameterization
are close to the experimental data for $\slope$ at fixed $Q^2$ and at
fixed
$x$ values over the kinematic range of interest. This parameterization
does not include a specific soft component. 
\item\,\,\,The GLMN model \cite{GLMNslope}, 
which is a pure pQCD dipole model with SC.
As such, the  model relates indirectly to gluon saturation, even though it
is constructed so as to include unitarity corrections below actual
saturation. This parameterization has no explicit soft component. 
\item\,\,\,The DL two Pomeron parameterization \cite{DL2P}. 

\end{enumerate}

All three models, as well as the BH parameterization, follow the
experimental behaviour of $\slope$ at fixed W rather well, including the
observed turnover. We demonstrate this in Fig.~5 by comparing the ZEUS
data \cite{ZEUSslope} with the results of our SC model. 
In Fig.~6 we plot the DL model
predictions versus the data at fixed W. We also display in this figure the
contribution of the hard DL Pomeron.

We conclude with three general comments: 

\begin{enumerate}
\item\,\,\,Gluon saturation 
is not unique in producing a turnover at fixed $W$.
\item\,\,\,The gluon saturation scale may be estimated
theoretically  from the contours produced at the boundary of
$\kappa_g\,=\,1$, as discussed in our papers
\cite{GLMslope,GLMNslope}. This theoretical analysis suggests that
gluon saturation occurs at $Q^2\, \approx $ 1 $GeV^2$. 
Note that $\kappa_g$ depends on $xG(x,Q^2)$, which is determined from a
DGLAP pdf fit to the global $F_2$ data.   
\item\,\,\,We note that in the BH and DL parameterizations 
the soft contributions are concentrated at a rather large
typical scale $\geq \,\,2 \,GeV^2$. This observation supports the
idea that the soft Pomeron originates from rather short
distances \cite{KL}.

\end{enumerate}

\begin{figure}
\centerline{\epsfig{file=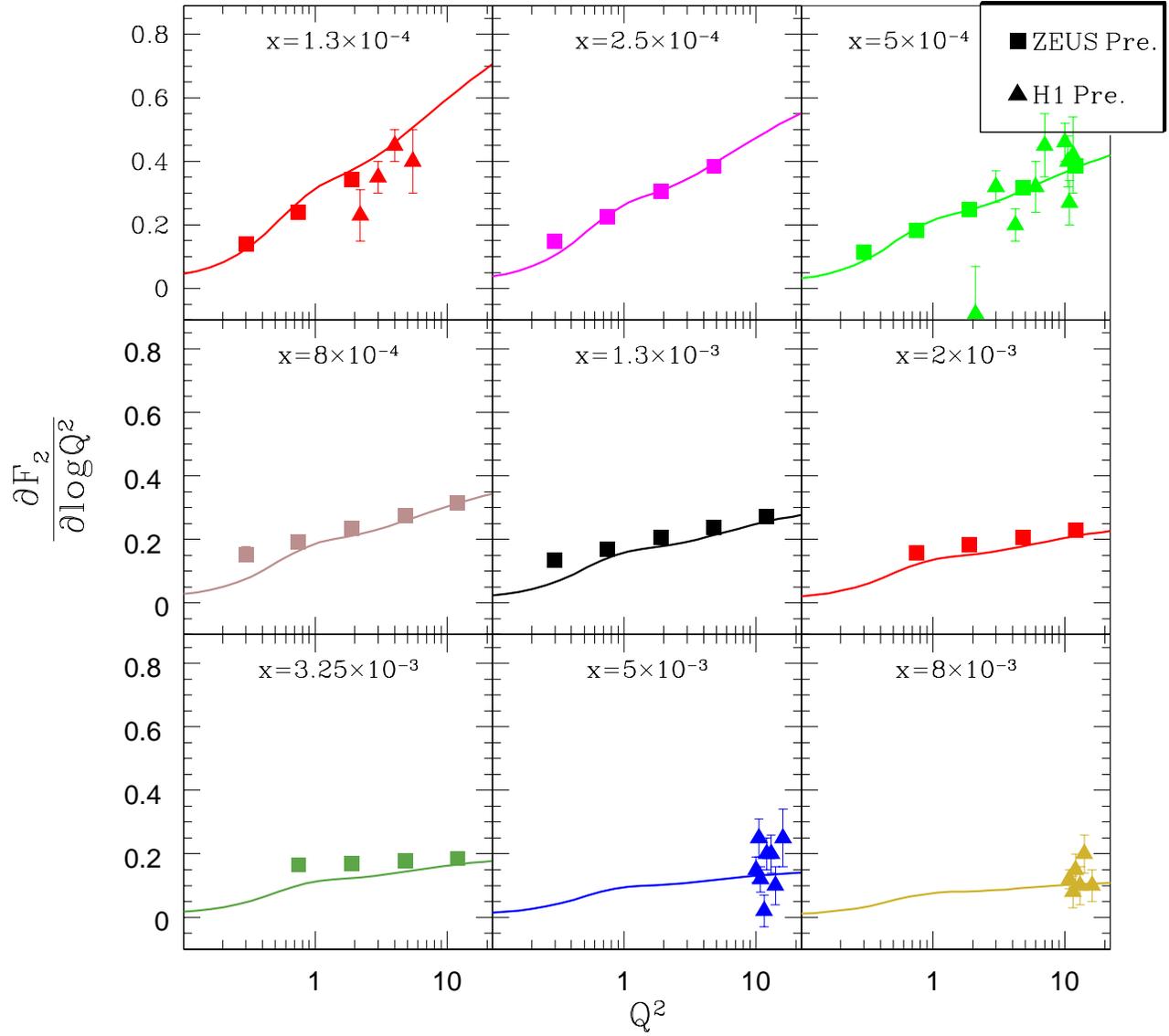 ,width=170mm}}
\caption{\it $Q^2$ (in $GeV^2$)  dependence of H1 and ZEUS logarithmic
slope data at
fixed $x$ compared with our calculations for screened GRV98.}
\label{Fig.3}
\end{figure}

\begin{figure}
\centerline{\epsfig{file=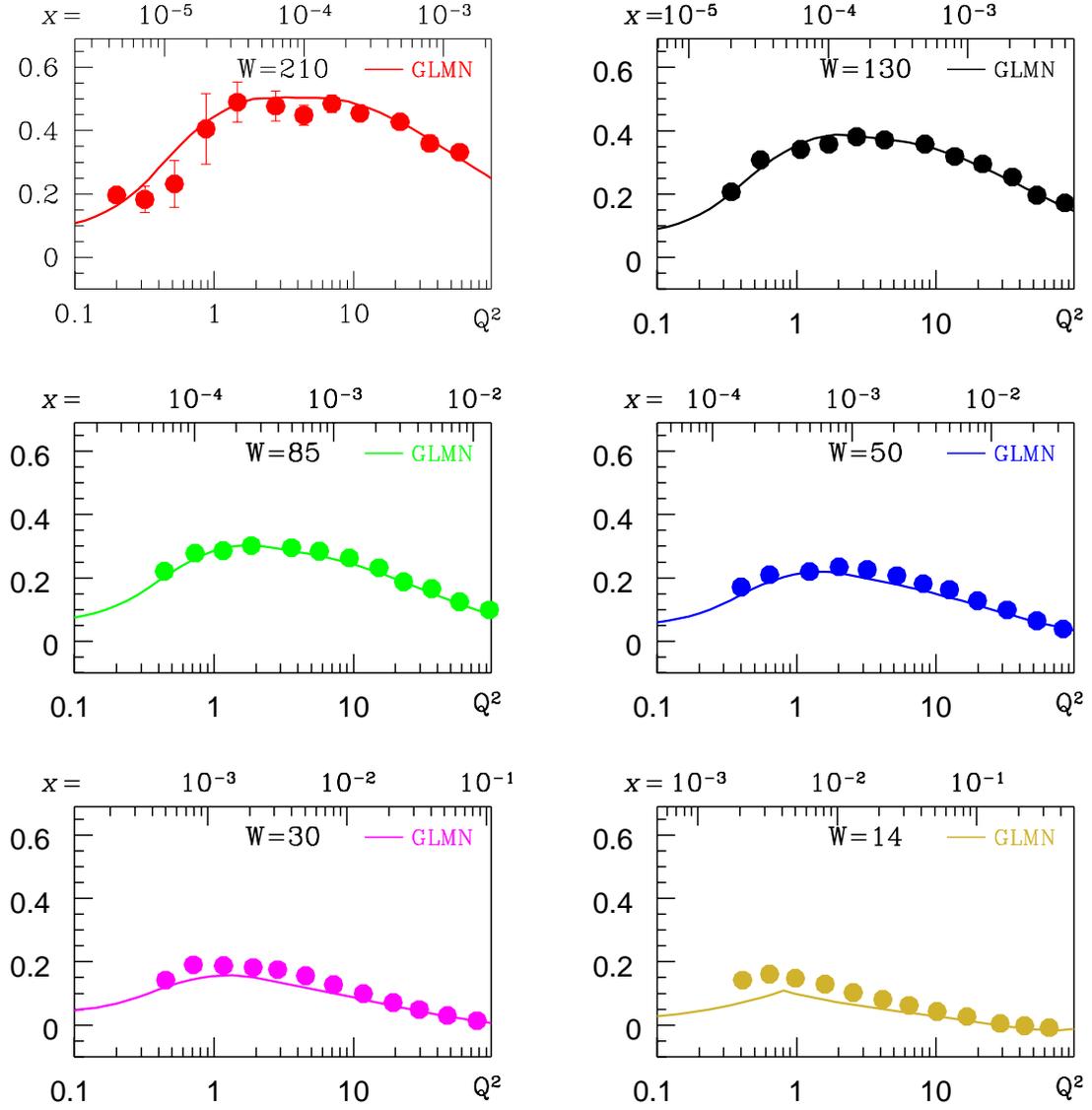,width=170mm}}
\caption{\it ZEUS logarithmic slope data at fixed W (in $GeV$)  compared
with our SC
calculation with  GRV98 input ( $Q^2$ in $GeV^2$).}
\label{Fig.4}
\end{figure}


\begin{figure}
\centerline{\epsfig{file=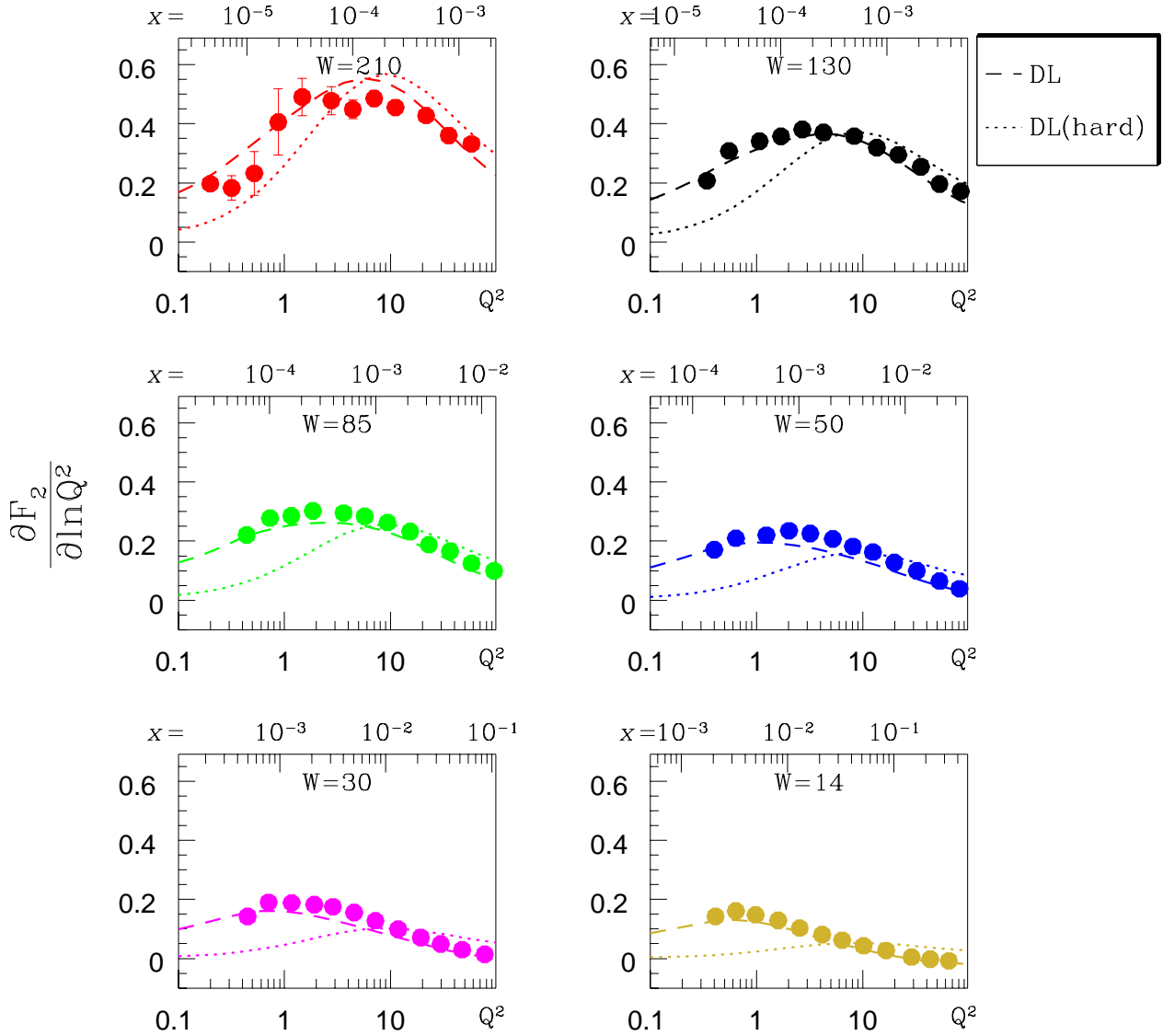,width=170mm}}
\caption{\it ZEUS logarithmic slope data at fixed W (in $GeV$)  compared
with DL two
Pomeron model ($Q^2$ in $GeV^2$).}
\label{Fig.5}
\end{figure}


\begin{figure}
\centerline{\epsfig{file=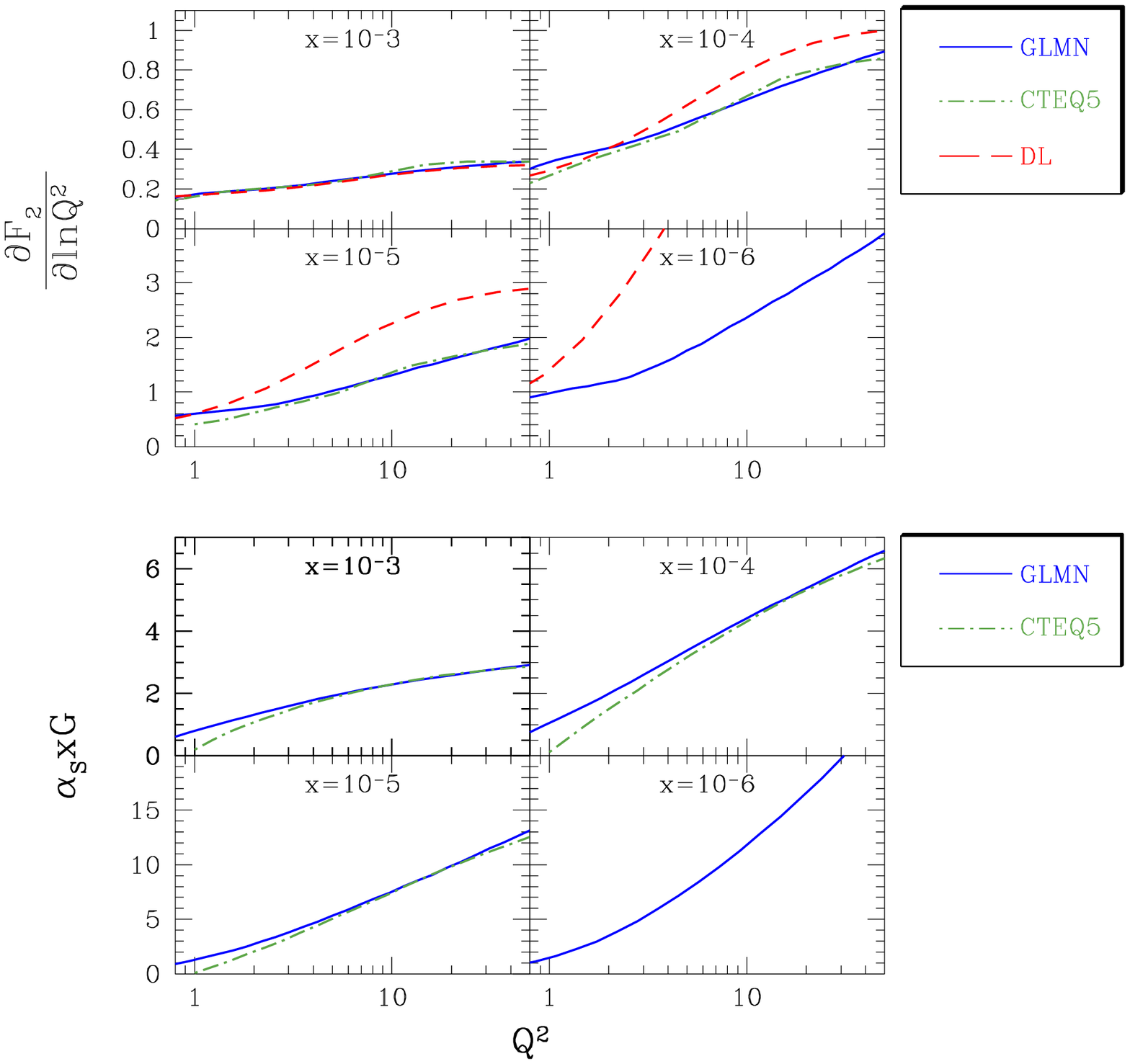,width=170mm}}
\caption{\it  Predictions for the different parameterizations 
 at fixed low values of x ($Q^2$ in $GeV^2$).}
\label{Fig.6}
\end{figure}

\section{Discussion and Conclusions}

The main experimental observation regarding $\slope$ is that 
at fixed $x$ it is a monotonic decreasing function of $Q^2$, and  
for fixed  $Q^2$ it increases as $x$ becomes smaller. This is the
observed pattern even for the lowest measured value of
$Q^2$ = ~0.3 $GeV^2$ \cite{ZEUSslope}.
The main objective of our investigation was to check whether the
preliminary H1 \cite {H1slope} and ZEUS \cite{ZEUSslope} data contains a
decisive experimental signature indicating signs of gluon
saturation. This effect is associated with the transition from the
relatively well understood pQCD hard domain to the more complex, and less
understood soft domain. We have  shown in this paper that the gluon
saturation is capable of describing all HERA  data on $\slope$ and,
therefore, the 
data do not contradict the idea that HERA has reached a new QCD regime: 
high parton density QCD.

However, the final conclusion of our analysis is that the present data can
be described by a number of models, which are very different in their
basic construction. Specifically, we show that all available data
with $Q^2 \, > \, Q_{0}^{2}$, of $\slope$
can be described by DGLAP using the CTEQ5 pdf as input, as
well as GRV98 pdf which have been corrected for screening. The DL
model which is Regge motivated and includes a significant soft component,
also  provides a good description of the data.

We have shown that the turnover seen in $\slope$ at fixed $W$, as a
function of either fixed $Q^2$ or $x$, is not necessarily connected with
saturation
effects and cannot be used to discriminate between models.
There has also been a successful reproduction of
the Caldwell plot by
Kaidalov et al. \cite{KMP} within the framework of the CKMT model
\cite{CKMT},
and by Forshaw et al.\cite{Forshaw} using a colour dipole model
formalism. We are of the opinion that fitting the Caldwell as well as the
fixed W  plots is a
necessary first step, but a comparison with all the detailed measurements 
is essential to
test the adequacy of a given model.

In Fig.~7 we attempt to clarify the similarities and differences
between the non screened CTEQ5 and the screened GRV98 DGLAP calculations
whose results are very similar, (see Figs.~2 and ~3). 
To this end we display the calculated distributions of $\slope$
and $\alpha_s xG(x,Q^2)$ at fixed values 
 $x \, =\, 10^{-3}, 10^{-4}, 10^{-5}$ and $10^{-6}$. As can be seen 
the results using CTEQ5 (non
screened) and GRV98 (screened)  in the limit of very small $x$  are very
similar. It is therefore suggestive,  that CTEQ5
contains significant
screening effects, that are absent in the boundary conditions used in
GRV98. Note that CTEQ5 is not defined for $ x\, < \, 10^{-5}$.  
Our estimates of SC for the CTEQ5 parameterization 
(based on  \eq{B} - \eq{D2}) is about 10\%, which should be considered as
the uncertainty in this  parameterization.
We also show the  DL  predictions for $\slope$, which differ from the
hard partonic DGLAP at small enough $x$.

The ZEUS data at exceedingly low $Q^2$ \cite{ZEUSlowQ2} are of
particular interest when investigating the transition from the DGLAP
dominated region to the non perturbative (low $Q^2$) region.
This transition was expected to be observed experimentally
since $F_2 \, \approx \, Q^2$ as $Q^2 \, \rightarrow \, 0 $ due to   
gauge invariance requirements, and for large $Q^2$, $F_2 \, \approx
\, \,( Q^2)^{\gamma}$ from DGLAP evolution, where $\gamma$ is the
anomalous dimension.
This transition has been seen by ZEUS
\cite{ZEUSlowQ2} in their measurements of $F_{2}(x,Q^2)$ at  
small values of $x$ and $Q^2$, where the transition appears to be at
$Q^2 \approx \, 1 \,\, GeV^2$, and is compatible with our early
theoretical estimates \cite{GLMslope,GLMNslope}. This provides an
indirect
indication that this is the scale for the onset of gluon saturation, but
we still lack a decisive signature for this effect.

Based on our present investigation we conclude that the 
behaviour of $\slope$ as measured in the kinematic region presently
accessible at  HERA, does not display  unambiguous signs of
saturation. This is compatible with the information displayed in Fig.~1
where we see that the anticipated gluon saturation scale is close to
$Q^2_0$, and to the scale at which the soft contribution becomes
significant.

We are extending our  search for possible effects of
gluon saturation to the channel of photo and DIS production of
$J/\Psi$ \cite{GFLMNpsi}, where the cross section should be very sensitive
to this signal, as it is proportional to $[xG(Q^2,x)]^2$.


{\bf Acknowledgments:}

UM wishes to thank UFRJ and FAPERJ (Brazil) for their support. EG and
EL would like to acknowledge the support and hospitality of the Theory
Group at DESY. EN would like to thank H. Abramowicz for a helpful
discussion.  This research was supported by in part by the Israel
Academy of Science and Humanities and by BSF grant \# 98000276.

\end{document}